# Fiber-Optic Interferometry Using Narrowband Light Source and Electrical Spectrum Analyzer: Influence on Brillouin Measurement

Yosuke Mizuno, *Member, IEEE*, Neisei Hayashi, and Kentaro Nakamura, *Member, IEEE*

*Abstract*— We observe an interference pattern using a simple fiber-optic interferometer consisting of an electrical spectrum analyzer and a narrowband light source, which is commonly employed for observing the Brillouin gain spectrum. This interference pattern expands well beyond the frequency range corresponding to the Brillouin frequency shift in silica fibers (~11 GHz at 1.55 µm). Using both silica single-mode and polymer optical sensing fibers, we then experimentally prove that the distinctive noise in a self-heterodyne-based Brillouin measurement with an unoptimized polarization state originates from the interference between the reference light and the Fresnel-reflected light. This noise can be almost completely suppressed by employing a delay line that is longer than the coherence length of the light source and by artificially applying a high loss near the open end of the sensing fiber.

## I. INTRODUCTION

SUBSTANTIAL efforts have been directed toward the study of Brillouin scattering in optical fibers [1] in the last 40 years, and a number of related applications have been developed, including lasing [1], microwave signal processing [2], core alignment [3], optical memory [4], slow light generation [5], and distributed strain and temperature sensing [6]–[10]. To improve the performance of these applications, Brillouin scattering properties have been investigated not only in standard silica glass fibers [11], [12] but also in various specialty fibers, such as tellurite fibers [13], [14], bismuth-oxide fibers [14], chalcogenide fibers [15], photonic crystal fibers [16], and fibers doped with rare-earth ions (erbium, thulium, etc.) [17], [18]. Each fiber type has its own distinctive features; for instance, the Brillouin scattering power in tellurite and chalcogenide fibers is far higher than that in other fibers, whereas the Brillouin scattering power in erbium-doped fibers at 1.55 µm can be tuned by controlling the 980-nm pump power. However, all of these glass fibers are fragile and require careful handling; in sensing applications, they cannot be used to measure strains larger than roughly 3%.

Our approach to overcome these problems exploits Brillouin scattering in polymer (or plastic) optical fibers (POFs) [19], [20], which can offer extremely high flexibility and can endure ~100% strain [21]. Though we have successfully observed and characterized Brillouin scattering in POFs [22]–[28], the power is quite low because of the large core, multimode nature, and relatively high propagation loss. For more detailed characterization, Brillouin measurements should be performed with a maximal signal-to-noise (SN) ratio.

In general, Brillouin scattering can be classified into two types: spontaneous scattering (with lower scattering power) and stimulated scattering (with higher scattering power). To date, stimulated Brillouin scattering (SBS) has been detected by the so-called pump-probe technique [29]–[32], which measures the change in scattering power as a function of the frequency difference between the pump and probe lightwaves. Because lock-in detection can be employed relatively easily, the SN ratio of the SBS measurement is sufficiently high even in POFs [25]. However, in measuring spontaneous Brillouin scattering (SpBS) with standard self-heterodyne detection, lock-in detection cannot be easily applied. As a result, the peculiar structure of the noise floor overlaps the Brillouin gain spectrum (BGS) and prevents the correct detection of the BGS and/or the Brillouin frequency shift (BFS). By compensating the noise floor of an electrical spectrum analyzer (ESA), the SN ratio was improved to some extent [33]. However, the SpBS measurement still suffers from this noise floor structure, especially when the polarization state is not precisely optimized. Clarifying the origin of this noise will enhance not only the SN ratio of the SpBS measurement (especially in POFs) but also the performance of self-heterodyne-based Brillouin sensors, such as Brillouin optical correlation-domain reflectometry (BOCDR) [10], [13], [33], [34].

In this study, we observe an interference pattern with a simple Mach-Zehnder interferometer consisting of an ESA and a narrowband light source, which is generally employed to observe the BGS, and show that the interference pattern expands the frequency range corresponding to the BFS in silica single-mode fibers (SMFs; ~11 GHz at 1.55 µm). We then experimentally prove, using a silica SMF and a POF as the fiber

Manuscript received *** **, 2014; revised *** ** 2014; accepted *** ** 2014. Date of publication *** **, 2014; date of current version *** **, 2014. This work was in part supported by Grants-in-Aid for Young Scientists (A) (no. 25709032) and for Challenging Exploratory Research (no. 26630180) from the Japan Society for the Promotion of Science (JSPS) and by research grants from the General Sekiyu Foundation, the Iwatani Naoji Foundation, and the SCAT Foundation. The work of N. Hayashi was supported by a Grant-in-Aid for JSPS Fellows (no. 25007652).

The authors are with Precision and Intelligence Laboratory, Tokyo Institute of Technology, Yokohama 226-8503, Japan (e-mail: ymizuno@sonic.pi.titech. ac.jp, hayashi@sonic.pi.titech.ac.jp, knakamur@sonic.pi.titech.ac.jp).

Color versions of one or more of the figures in this paper are available online at http://ieeexplore.ieee.org.

Digital Object Identifier **.****/JLT.2014.******

under test (FUT), that the aforementioned distinctive noise in the SpBS measurement with an unoptimized polarization state originates from interference between the reference light and Fresnel-reflected light. We also show that this noise can be almost completely suppressed by employing a delay line that is longer than the coherence length of the light source and by artificially applying a considerable loss near the open end of the FUT.

## II. PRINCIPLES

### A. Brillouin Scattering

Spontaneous Brillouin scattering in optical fibers is caused by acoustic-optical interaction, generating backscattered Stokes light, the spectrum of which is referred to as the BGS [1]. The central frequency of the BGS shifts downward relative to the incident pump frequency by an amount termed as BFS. At 1.55 μm, the BFS is reported as ~10.8 GHz for silica SMFs [1] and ~2.8 GHz for perfluorinated graded-index POFs [22]. The BFS depends on strain and temperature [11], [12], [23], [26], which is the basic operating principle of Brillouin-based distributed sensors, such as Brillouin optical time-, frequency-, and correlation-domain analysis (BOTDA [6], [35], [36], BOFDA [8], [37], [38], BOCDA [9], [39], [40], respectively) and Brillouin optical time- and correlation-domain reflectometry (BOTDR [7], [41], [42] and BOCDR [10], [13], [33], [34], respectively).

Because the BFS values in optical fibers are not sufficiently smaller than the frequency resolution of a typical optical spectrum analyzer (OSA; ~10 GHz), an ESA is often employed with a self-heterodyne scheme to observe the BGS with a much higher frequency resolution. However, self-heterodyne detection is known to broaden the observed BGS by double the amount of the 3-dB linewidth of the pump light [43], leading to possible incorrect measurements. Therefore, a light source with an output linewidth sufficiently narrower than the linewidth of the BGS should be utilized. The Brillouin linewidth is typically ~30 MHz in silica SMFs [1] and ~100 MHz in POFs [22] at 1.55 μm (depending on the pump power [24], [43]). Therefore, a laser diode (LD) with ≤ 1-MHz linewidth is suitable. Moreover, the Brillouin scattering power measured by self-heterodyne detection is highly dependent on the relative polarization state between the Stokes light and reference light, which sometimes leads to unstable measurements [33], [44], [45].

### B. Fiber-Optic Interferometry

Fiber-optic interferometry can be used as a self-heterodyne detection system for SpBS measurement. In general, when the output of a broadband light source is injected into an optical interferometer, its transmission spectrum exhibits an interference pattern with periodic peaks and dips. Suppose the interferometer is composed of one type of fiber. Then, the period of the interference pattern $\Delta f$ is inversely proportional to the difference between the two path lengths $\Delta L$ as

$$\Delta f = \frac{c}{n\,\Delta L}, \qquad (1)$$

where $c$ is the velocity of light in vacuum and $n$ is the refractive index (of the fiber core in this case). Therefore, by measuring $\Delta f$, the change in either refractive index or path length can be derived. To date, a number of studies have been published relating to sensing based on this principle [46]–[50]. However, these studies commonly used broadband light sources (such as super continuum sources, amplified spontaneous emission outputs, super luminescent light emitting diodes, and swept-source lasers) and OSAs. No reports have been provided on the use of narrowband light sources and ESAs.

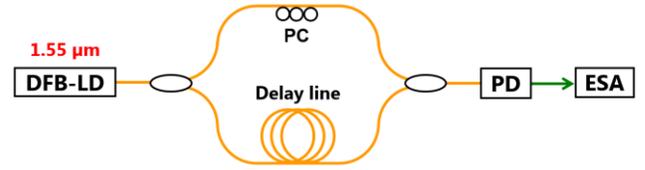

Fig. 1. Schematic of Mach-Zehnder interferometer used in the experiment, containing the distributed-feedback laser diode (DFB-LD), electrical spectrum analyzer (ESA), polarization controller (PC), and photo diode (PD).

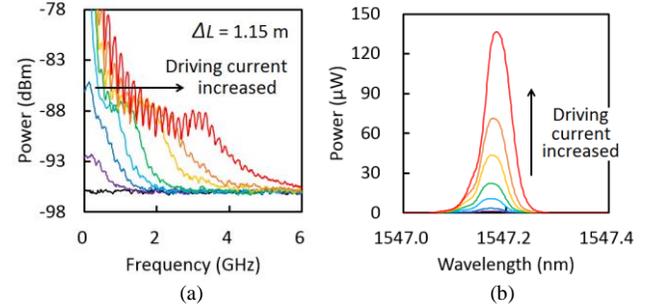

Fig. 2. (a) Observed expansion of the interference pattern. Each electrical spectrum was measured at a driving current of 11.0 (black), 11.1 (purple), 11.2 (dark blue), 11.4 (light blue), 11.9 (green), 12.7 (yellow), 13.8 (orange), and 16.2 mA (red). (b) Corresponding optical spectra of the LD output.

## III. EXPERIMENTS

### A. Mach-Zehnder interferometry

To verify the operation of the fiber-optic interferometer using a narrowband light source and an ESA, we implemented a standard Mach-Zehnder interferometer (Fig. 1). We used a distributed-feedback LD (NX8562LB, NEC; 1.55 μm; 10 mW output) with a 3-dB linewidth of 1 MHz, corresponding to a coherence length of approximately 200 m. The laser output was divided and guided into two arms; a polarization controller (PC) was inserted in one arm to control the relative polarization state, while a delay line was inserted in the other arm to control the path-length difference. The two beams were then coupled, converted into an electrical signal with a photo diode (PD), and observed with an ESA.

Figure 2(a) shows the measured electrical spectra when the LD driving current was gradually increased around its threshold (~11 mA). The path-length difference $\Delta L$ was 1.15 m. The interference pattern expanded to a larger frequency range

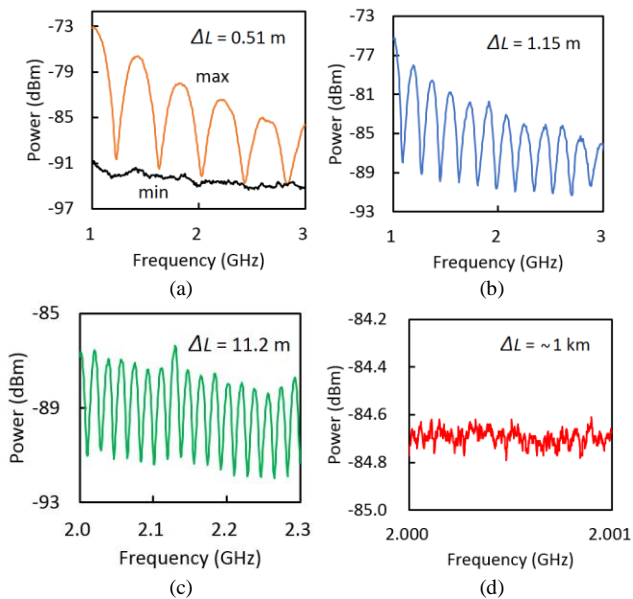

Fig. 3. Measured interference patterns around 2 GHz when the path-length difference was (a) 0.51 m, (b) 1.15 m, (c) 11.2 m, and (d) ~1 km. The polarization state was adjusted to maximize the visibility. The additional trend in (a) indicates the spectrum when the polarization state was adjusted to minimize the visibility.

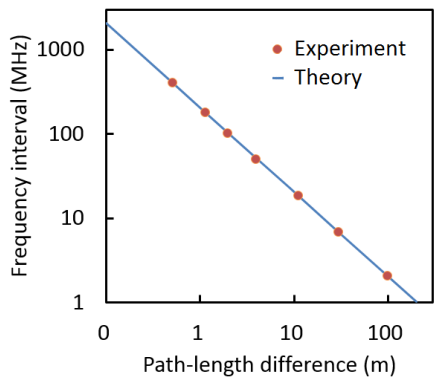

Fig. 4. Frequency interval plotted as a function of the path-length difference. The circles are measured data, and the blue line is a theoretical line calculated using Eq. (1).

as the LD output of the optical spectrum increased, as shown in Fig. 2(b).

Figures 3(a)–(d) show the measured interference patterns around 2 GHz when the path-length difference $\Delta L$ was 0.51 m, 1.15 m, 11.2 m, and ~1 km, respectively. The polarization state was adjusted to maximize the interferometric visibility (or amplitude of the interference pattern). The frequency intervals of the interference patterns were 401.1 MHz ($\Delta L = 0.51$ m), 178.6 MHz ($\Delta L = 1.15$ m), and 18.2 MHz ($\Delta L = 11.2$ m). The theoretical interval for $\Delta L = $ ~1 km, which is ~0.2 MHz, was not observed regardless of the polarization state, because $\Delta L$ was much longer than the coherence length (~200 m). Figure 3(a) also displays the electrical spectrum measured when the polarization state was adjusted to minimize the visibility. The pattern was no longer observed. Thus, the interference pattern was almost completely suppressed by optimizing the polarization state; however, the emergence of the interference pattern cannot be avoided in many practical cases, such as cases where adaptive polarization control cannot be performed or where polarization diversity [44] or polarization scrambling [33], [45] is employed.

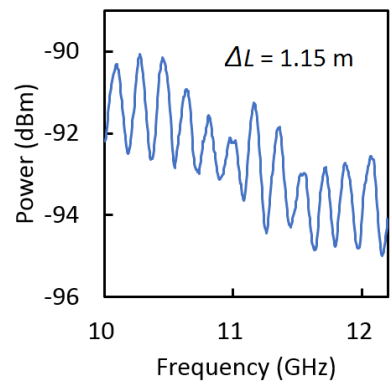

Fig. 5. Measured interference pattern around 11 GHz when the path-length difference was 1.15 m. The polarization state was adjusted to maximize the visibility.

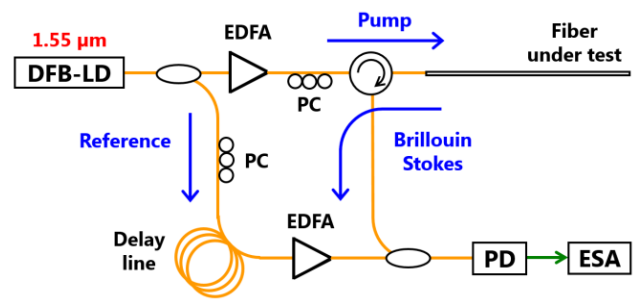

Fig. 6. Schematic of the self-heterodyne-based experimental setup for Brillouin measurement, containing the distributed-feedback laser diode (DFB-LD), erbium-doped fiber amplifiers (EDFAs), electrical spectrum analyzer (ESA), polarization controllers (PCs), and photo diode (PD).

The relationship between the frequency interval and the path-length difference was in extremely good agreement with the theoretical calculations (Fig. 4). This result implies that the basic operation of fiber-optic interferometry (such as the measurement of path-length difference) can be achieved in the spectral domain by using a narrowband light source (including a cost-effective DFB-LD) and an ESA. The applicable frequencies range from several megahertz (corresponding to the coherence length of the light source) to several gigahertz (corresponding to the upper limit of the measurable frequency of a standard ESA). This method is free from errors caused by power fluctuations, leading to higher stability than that of the well-known power-domain fringe-counting techniques using a narrowband light source and an optical power meter (or an oscilloscope) [51]–[53].

Figure 5 shows the interference pattern near 11 GHz for $\Delta L = 1.15$ m. The interference pattern expands well beyond the range where the BGS in a standard silica SMF is observed at 1.55 μm. Therefore, this interference pattern should have a significant effect on the Brillouin measurement when the polarization state is not optimized.

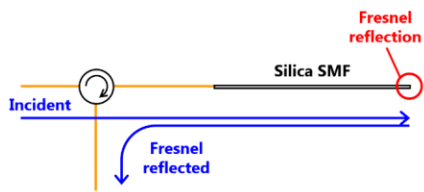

Fig. 7. Definition of the "pump path" when a silica SMF was used as the FUT.

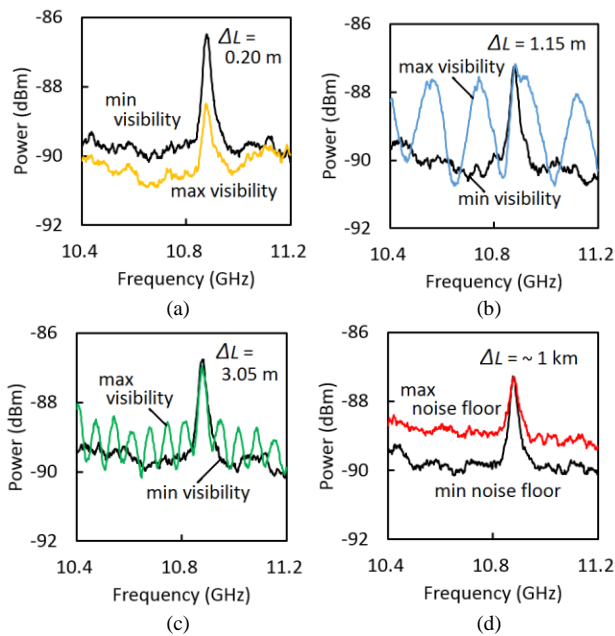

Fig. 8. Measured electrical spectra around 10.8 GHz when the path-length difference was (a) 0.20 m, (b) 1.15 m, (c) 3.05 m, and (d) ~1 km. In (a)–(c), the polarization state was adjusted to maximize (colored) and minimize (black) the interferometric visibility. In (d), the polarization state was adjusted for the highest (colored) and lowest (black) noise floor.

### B. Brillouin measurement in silica SMFs

First, we investigated the influence of the interference pattern on the Brillouin measurement using a 13.2-m-long silica SMF as the FUT. The experimental setup used in the experiment, which is based on self-heterodyne detection, is depicted in Fig. 6. All optical paths were composed of silica SMFs. The DFB-LD output (same as that described in the preceding section) was divided into two beams: one was used as pump light, and the other was used as reference light. The pump light was amplified to 20 dBm with an erbium-doped fiber amplifier (EDFA) and injected into the FUT, generating backscattered Stokes light. The reference light was amplified to 3 dBm with the EDFA and coupled with the Stokes light. The optical beat signal was converted into an electrical signal and observed with an ESA as BGS. The relative polarization state between the Stokes light and the reference light was adjusted with two PCs. A delay line was inserted in the reference path to control the path-length difference $\Delta L$ between the pump path and reference path. Here, we define the "pump path" as the path along which the light Fresnel-reflected at the end of the FUT propagates (see Fig. 7).

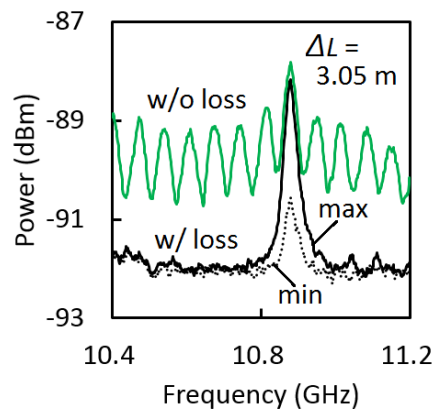

Fig. 9. Measured electrical spectra around 10.8 GHz when the path-length difference was 3.05 m before (green) and after (black) the bending loss was applied near the FUT end. The polarization state was adjusted to maximize (solid) and minimize (dotted) the Brillouin peak power.

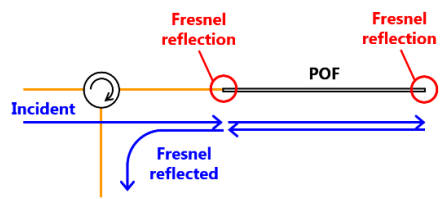

Fig. 10. Two possible pump paths when a POF was used as the FUT.

Figures 8(a)–(d) show the measured electrical spectra around 10.8 GHz when the path-length difference $\Delta L$ was 0.20 m, 1.15 m, 3.05 m, and ~1 km, respectively. The spectral peak observed at 10.87 GHz is the BGS in the silica SMF [1]. In Figs. 8(a)–(c), the polarization state was adjusted to maximize and minimize the interferometric visibility. In Fig. 8(d), the polarization state was adjusted for the highest and lowest noise floor because the height of the noise floor is positively correlated with the visibility in this experiment. In Fig. 8(a), the BGS was detected even though it overlapped the interference pattern (its dip can be observed at ~10.65 GHz) because the theoretical frequency interval of ~1.0 GHz was much broader than the Brillouin linewidth (though the SN ratio was lower). In Figs. 8(b) and (c), the frequency intervals were 178.3 MHz and 67.3 MHz, respectively, overlapping the BGS and resulting in erroneous BGS measurement. These interference patterns were drastically mitigated by properly adjusting the polarization state. In Fig. 8(d), as ~1 km was much longer than the coherence length of the LD, the interference pattern was not observed regardless of the polarization state.

One method for suppressing the influence of the interference pattern is to weaken the Fresnel reflection at the FUT end. Figure 9 shows the measured electrical spectra for $\Delta L = 3.05$ m before and after a bending loss (> 60 dB) was applied near the FUT end. Using this method, the interference pattern was almost completely diminished, and the polarization state adjustment was the sole contribution to the Brillouin peak power, which is an ideal situation for Brillouin measurement.

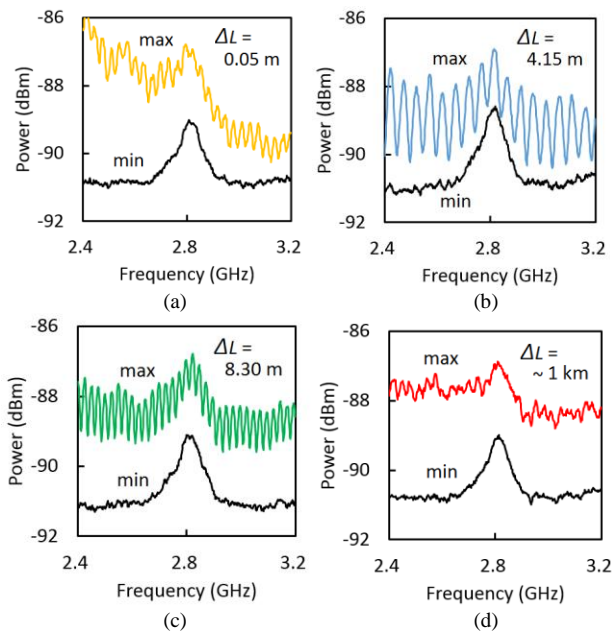

Fig. 11. Measured electrical spectra around 2.8 GHz when the path-length difference was (a) 0.05 m, (b) 4.15 m, (c) 8.30 m, and (d) ~1 km. The polarization state was adjusted so that the visibility was maximized (colored) and minimized (black).

### C. Brillouin measurement in POFs

Next, we performed similar experiments using POF as the FUT. The same experimental setup (Fig. 6) was employed. The pump power was amplified to 25 dBm, and the reference power was 3 dBm. The POF used in the experiment was a 3.57-m-long perfluorinated graded-index POF with a core diameter of 50 μm, a core refractive index of 1.35, an outer diameter of 500 μm, a numerical aperture of 0.185, and a propagation loss of ~250 dB/km at 1.55 μm. Both ends of the POF were cut perpendicular to the fiber axis, and one end was butt-coupled [22] to the pigtail (silica SMF) of the optical circulator. As shown in Fig. 10, Fresnel reflection occurs at two points: the SMF/POF interface and the open end of the POF. Considering that the light, which is Fresnel-reflected at the open end, drastically attenuates when it returns from the POF to the silica SMF because of the serious mode-field diameter mismatch [22], we define the main "pump path" (used to calculate the path-length difference $\Delta L$) as the path along which the Fresnel-reflected light propagates at the SMF/POF interface.

Figures 11(a)–(d) show the measured electrical spectra around 2.8 GHz when $\Delta L$ was 0.05 m, 4.15 m, 8.30 m, and ~1 km, respectively. The spectral peak observed at 2.81 GHz is the BGS in the POF [22]. The polarization state was adjusted to maximize and minimize the visibility. In Figs. 11(b) and (c), the measured frequency intervals were 49.3 MHz and 24.4 MHz, respectively, which overlapped the BGS in a manner analogous to the silica SMF case. However, unlike the silica SMF case, weak interference patterns were observed even for $\Delta L = 0.05$ m (Fig. 11(a); the theoretical frequency interval is ~4.1 GHz) and $\Delta L = $ ~1 km (Fig. 11(d); beyond the coherence length). The measured frequency intervals for both path lengths were ~31.8 MHz, which corresponds to two lengths of the 3.57-m-long POF ($n = 1.35$). Therefore, we conclude that these interference

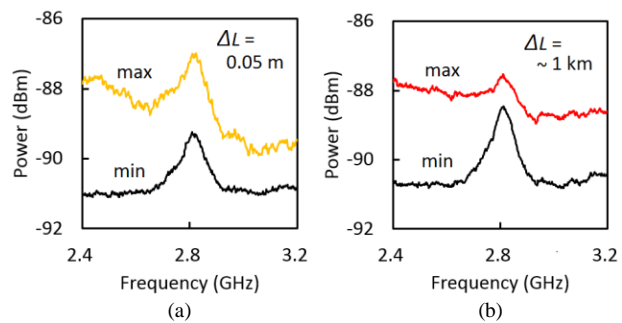

Fig. 12. Measured electrical spectra when $\Delta L$ was (a) 0.05 m and (b) ~1 km, after the open end of the POF was angled-cut. The polarization state was adjusted so that the noise floor was highest (colored) and lowest (black).

patterns were caused by the interference between two pump waves Fresnel-reflected at the SMF/POF interface and at the open end of the POF.

Subsequently, we cut the open end of the POF at an angle to suppress the Fresnel reflection (note that applying a high bending loss to POFs is difficult [20]), and we measured the electrical spectra when $\Delta L$ was 0.05 m and ~1 km, as shown in Figs. 12(a) and (b), respectively. The polarization state was adjusted for the highest and lowest noise floor. Even when the noise floor was high, the interference patterns were greatly suppressed. The polarization state adjustment was not influenced by the interference pattern. However, to maximize the SN ratio of the BGS measurement (i.e., the difference between the Brillouin peak power and the noise floor), the suppression of the Rayleigh noise (tail of the Rayleigh-scattered light spectrum) is more important than the enhancement of Brillouin peak power [54]. Note that Rayleigh noise is a phenomenon unique to POFs with a low BFS. As the polarization state for maximal Brillouin peak power is generally different from that for minimal Rayleigh noise, the optimized BGS with the highest SN ratio in Fig. 12(a) differs from that in Fig. 12(b).

### D. Discussion

When the polarization state cannot be adaptively optimized, the interference pattern emerges as a serious problem in observing the BGS. In the case of silica SMFs, the interference pattern can be mitigated either by adjusting the fiber length difference between the pump and reference paths or by applying a high bending loss near the SMF end (or by cutting the end at an angle). When the former method is used, a delay line, which is longer than the coherence length of the light source, should be inserted into one of the paths because shortening the path-length difference requires considerable effort. In BOCDR, we often insert a long delay line in the reference path to control the order of the correlation peak [10], [34], which is also desirable from the aspect of SN ratio enhancement. On the other hand, in the case of POFs, the interference pattern cannot be sufficiently suppressed by adjusting the path-length difference or by suppressing the Fresnel reflection at the open end; these two measures must be taken simultaneously. The suppression of the Fresnel reflection at the open end can be achieved either by cutting the end at an angle, immersing the end into index-matching oil or water ($n = $ ~1.32 at 1.55 µm [55]), or using a sufficiently long POF to attenuate the Fresnel-reflected light.

For the BGS measurement, the polarization state adjustment has three roles: maximization of the Brillouin peak power, minimization of the Fresnel-induced interference pattern, and minimization of the Rayleigh noise (especially in the case of POFs). These three behaviors are independent of one another; for instance, a minimized interference pattern generally does not lead to a maximized Brillouin peak power. Therefore, by suppressing the influence of the interference pattern using one of the aforementioned methods, the polarization state can be adjusted solely to maximize the Brillouin peak power in silica SMFs or to minimize the Rayleigh noise in the case of POFs [54]. This interference pattern suppression results in BGS and BFS detection with the highest SN ratio, even when the polarization state cannot be arbitrarily adjusted.

## IV. CONCLUSION

We clarified the origin of the distinctive noise in SpBS measurements with an unoptimized polarization state and presented a strategy for detecting the BGS with the highest achievable SN ratio. First, using a Mach-Zehnder interferometer consisting of an ESA and a narrowband light source, we demonstrated an interference pattern that expanded well beyond the ~11 GHz range. Then, using a silica SMF and a POF as the FUT, the distinctive noise in the SpBS measurement was shown to originate from the interference between the reference light and the Fresnel-reflected light. Moreover, this noise is almost completely suppressed by inserting a delay line longer than the coherence length of the light source and by applying a considerable loss near the open end of the FUT. We believe that these results will be a significant guideline for characterizing the Brillouin properties in various fibers with a high SN ratio as well as for developing high-performance fiber-optic Brillouin devices and systems in the near future.

**Yosuke Mizuno** (M'14) was born in Hyogo, Japan, on October 13, 1982. He received the B.E., M.E., and Dr.Eng. degrees in electronic engineering from the University of Tokyo, Japan, in 2005, 2007, and 2010, respectively.

From 2007 to 2010, he was engaged in Brillouin optical correlation-domain reflectometry for his Dr.Eng. degree at the University of Tokyo. From 2007 to 2010, he was a Research Fellow (DC1) of the Japan Society for the Promotion of Science (JSPS). From 2010 to 2012, as a Research Fellow (PD) of JSPS, he worked on polymer optics at Tokyo Institute of Technology, Japan. In 2011, he stayed at BAM Federal Institute for Materials Research and Testing, Germany, as a Visiting Research Associate. Since 2012, he has been an Assistant Professor at the Precision and Intelligence Laboratory, Tokyo Institute of Technology, where he is active in fiber-optic sensing, polymer optics, and ultrasonics.

Dr. Mizuno is the winner of the Funai Research Award 2010, the Ando Incentive Prize for the Study of Electronics 2011, the NF Foundation R&D Encouragement Award 2012, and the Challenging Research Award 2013. He is a member of the IEEE, the Japanese Society of Applied Physics (JSAP), and the Institute of Electronics, Information, and Communication Engineers (IEICE) of Japan.

**Neisei Hayashi** was born in Gunma, Japan, on April 13, 1988. He received the B.E. degree in advanced production from the Gunma National College Technology (GNCT), Japan, in 2011, and the M.E. degree in electronic engineering from Tokyo Institute of Technology, Japan, in 2013.

From 2008 to 2010, he studied a ring-wandering phenomenon for his B.E. degree at the GNCT. From 2011 to 2013, he worked on nonlinear optics in polymers for his M.E. degree at Tokyo Institute of Technology. In 2013, he became a Research Fellow (DC1) of the Japan Society for the Promotion of Science (JSPS), and, since then, he has been continuing to study polymer optics for his Dr.Eng. degree at Tokyo Institute of Technology. His research interests include fiber-optic sensing, polymer optics, and ultrasonics.

Mr. Hayashi is the winner of the TELECOM System Technology Award for Student 2012. He is a member of the Japanese Society of Applied Physics (JSAP), and the Institute of Electronics, Information, and Communication Engineers (IEICE) of Japan.

**Kentaro Nakamura** (M'00) was born in Tokyo, Japan, on July 3, 1963. He received the B.E., M.E., and Dr.Eng. degrees from Tokyo Institute of Technology, Japan, in 1987, 1989, and 1992, respectively.

Since 2010, he has been a Professor at the Precision and Intelligence Laboratory, Tokyo Institute of Technology. His research field is the applications of ultrasonic waves, measurement of vibration and sound using optical methods, and fiber-optic sensing.

Prof. Nakamura is the winner of the Awaya Kiyoshi Award for encouragement of research from the Acoustical Society of Japan (ASJ) in 1996, and the Best Paper Awards from the Institute of Electronics, Information and Communication Engineers (IEICE) in 1998 and from the Symposium on Ultrasonic Electronics (USE) in 2007 and 2011. He also received the Japanese Journal of Applied Physics (JJAP) Editorial Contribution Award from the Japan Society of Applied Physics (JSAP) in 2007. He is a member of the IEEE, the ASJ, the JSAP, the IEICE, and the Institute of Electrical Engineers of Japan (IEEJ).